# An Empirical Assessment of the Accounting Semi-Identity Problem's Pervasiveness and Severity

F. Javier Sánchez-Vidal

Universidad Politécnica de Cartagena (Spain), Faculty of Business Studies, https://orcid.org/0000-0003-0587-2927

**Abstract** This paper investigates a fundamental methodological flaw in the investment-cash flow sensitivity model of Fazzari, Hubbard, and Petersen (1988). The model comes from a full accounting identity in which some components are missing, generating what I term an Accounting Semi-Identity, that mechanically links investment and cash flow, and this could bias coefficients, making the estimation difficult if not impossible. I propose an augmented specification including a variable that captures this arithmetic bias and test it across multiple firm-level databases. Results show that the ASI distortion is universal and severe: the ASI issue is present in 100% of the databases and explains more than 83% of the total explained variance, while the standard Fazzari, Hubbard, and Petersen (1988) model only accounts for approximately 17%. These findings suggest that a substantial body of prior empirical research based on this model may have reported spurious results rather than evidence of underlying economic behavior. This finding provides a compelling explanation for the substantial body of literature surrounding this model that has reported anomalous and incoherent results.

**Keywords:** accounting identities; accounting semi-identities; investment-cash flow sensitivity; open science, data reuse.

**JEL Classifications:** C13, C51, G31

## 1. Introduction

The model developed by Fazzari, Hubbard, and Petersen (1988) (hereafter FHP) relates investment, or increases in long-term assets, to cash flows in order to assess the degree of financial constraints faced by a firm. The estimated coefficient of the independent variable, cash flow, is referred to as the investment–cash flow sensitivity. For firms that are highly financially constrained, this coefficient is expected to be both large and statistically significant. But both variables are also related through a linear relation, as they are key elements of the accounting identity (AI) that represents the growth a balance sheet experiences from one period to the next. As some components of the full AI are missing, the FHP is an accounting semi-identity (ASI). This characteristic can pose problems for estimating its linear regressions, as the force of causality may be mixed with the force of arithmetic in the coefficient estimates, a point already emphasized by Sánchez-Vidal (2023). Using Monte Carlo simulations, this author demonstrates how these incomplete AIs can distort regressions, leading to biased coefficient estimates and inflated $R^2$ values.

The initial formulation of this methodological critique regrettably went largely unnoticed, a phenomenon mirroring the delayed reception of many scientific challenges. This lack of immediate engagement can be attributed to scientific inertia and inherent skepticism. Established methodologies create significant path dependencies, embedded in previous literature research, textbooks and peer-review norms, creating a high barrier to the adoption of new paradigm.

This pattern of delayed recognition is well-documented across disciplines. In economics, the profound critiques of macroeconomic models raised by Lucas (1976) in his "Lucas Critique" took years to be fully integrated, as the profession grappled with moving beyond structures that ignored agents' changing expectations. Similarly, in finance, early empirical anomalies challenging the Efficient Market Hypothesis (Fama, 1970), such as the work of De Bondt and Thaler (1985) on investor overreaction, were initially dismissed as statistical artifacts before the accumulation of evidence gave rise to the field of behavioral finance. In other fields, the replication crisis in psychology (Open Science Collaboration, 2015) was preceded by decades of unheeded warnings about questionable research practices and p-hacking (Cohen, 1994; Ioannidis, 2005), which went initially ignored. These examples illustrate a common trajectory: some critiques often face initial resistance not necessarily due to a lack of merit, but due to the cognitive and institutional friction that must be overcome.

While a degree of skepticism is understandable, researchers, particularly those who have employed the FHP (1988) model, are expected to maintain intrinsic scientific curiosity. The procedure for verifying the existence of an ASI problem, explained in Section 3 and already detailed in Sánchez-Vidal (2023), is methodologically straightforward. However, for reasons that remain unclear, even those who have presumably read the 2023 publication have, to the best of my knowledge, not attempted to test for the potential influence of the ASI on their results.

Reproducibility in Open Science is the practice of ensuring that scientific findings can be independently verified by others using the original data. The FHP (1988) model is an ideal candidate for data reuse and reproducibility analysis, as it has produced contradictory results for decades. leading to ongoing skepticism regarding the model's general applicability (Kaplan and Zingales, 1997, 2000; Machokoto et al., 2021; Adu-Ameyaw et al., 2022 and Wang, 2022, between others). To empirically assess whether the ASI problem and its inherent arithmetic forces influence the results of previously published articles working with the FHP (1988) model, I created a list of all articles that applied the FHP (1988) model, a well-known ASI-based model (Sánchez-Vidal (2023), published in journals listed in the Journal Citation Reports (JCR) between 2018 and the end of 2023.

The procedure was to contact the authors and to gently request their databases, and then determine the extent to which the ASI problem may have biased their published results and to evaluate the practical significance of the ASI critique. Of the 52 articles identified, I successfully obtained the datasets from only two. Notably, 19 of the 52 articles explicitly stated that data were "available upon request," or similar, yet these requests yielded no response. Consequently, I adopted an alternative approach by supplementing the two newly acquired datasets with seven from my prior work, resulting in a total of nine databases for the subsequent analysis. The empirical tests for the presence of the ASI problem, conducted across these nine datasets, are presented in Section 4.

Results show that the presence of the ASI problem is generalized, affecting 100% of the databases and confirming the hypothesis regarding the distorting influence of the ASI. The remainder of this paper is structured as follows. Section 2 elaborates on the ASI problem. Section 3 details the models, variables, and databases employed. Section 4 presents and discusses the results. Section 5 is a concise note on the open science and data reuse practices of this article, explaining how to access the data and reproduce its findings. Finally, Section 6 provides the concluding remarks.

## 2. The ASI problem

Sánchez-Vidal (2023) shows that using investment and cash flow to measure financial constraints can be misleading. This is because the two variables are mathematically related through an AI, a fact often overlooked. Both are components (and likely to be important) of both sides of the balance sheet taken in increases from one period to another. The following figure describes this full AI.

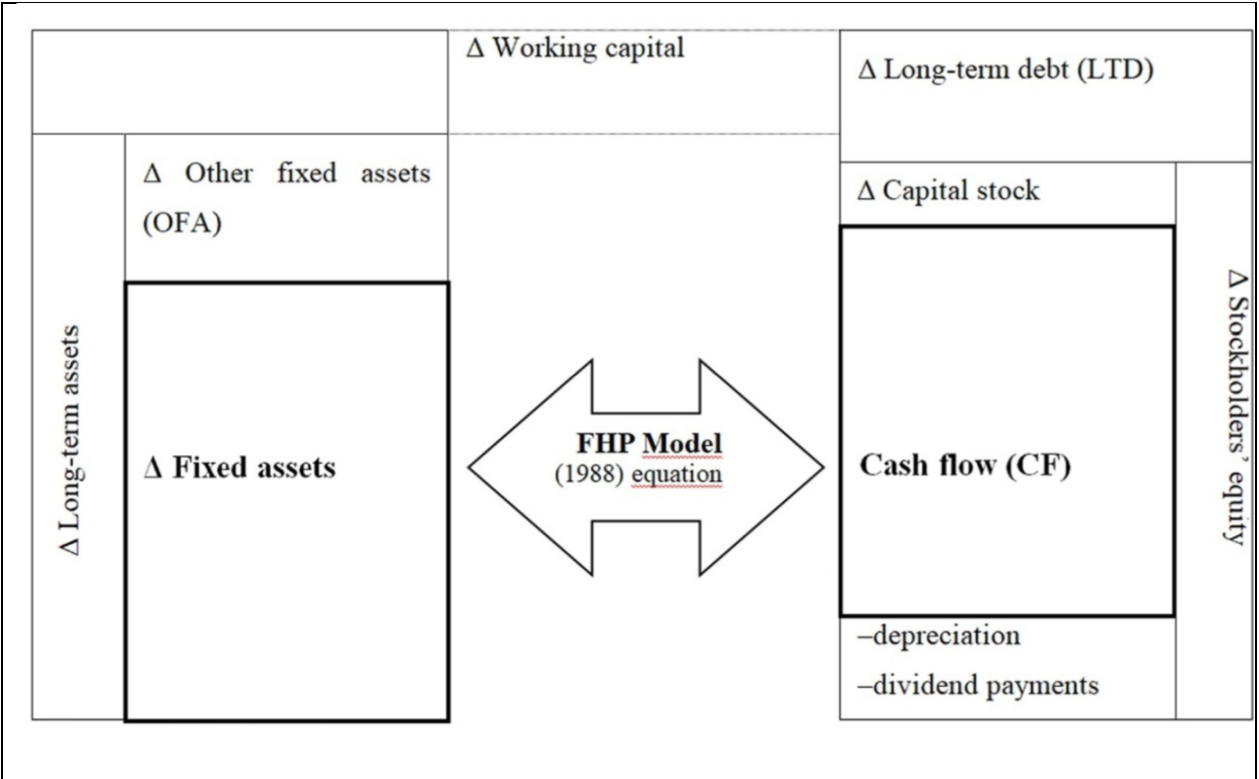

**Figure 1. Demonstration of the ASI in the FHP (1988) model.**
Source: Sanchez-Vidal, F.J. (2023)

The full AI is given by the expression:

$$Inv = cf + (\Delta LTD + \Delta Capital\ Stock - Deprec. - Dividends - \Delta Working\ capital - \Delta OFA) \quad (1)$$

Sanchez-Vidal (2023) calls all the elements in the parentheses the "rest,"

The model proposed by FHP (1988) is based on this regression:
$$Inv = a + b_1 * cf + \varepsilon \quad (2)$$

This specification omits the rest of the full AI, making it an ASI. Consequently, the results are potentially biased because the estimated coefficient ($b_1$) must mechanically compensate for this omission. As Farris et al. (1989) noted, in models derived from AIs, the coefficients of the included variables will adapt to account for any missing components.

The direction of this bias depends on the "rest." Sanchez-Vidal (2023) provides an illustrative example: a firm invests $5 million, generates $7 million in cash flow, and has a rest of -$2 million. The full AI would be: 5 = 7 - 2. However, as the "-2" is omitted from the regression, the cash flow coefficient must adjust downward to 0.714 to force the equation to hold (5 = 0.714 × 7). Following this logic, the value of the cash flow coefficient is mechanically determined by the rest:

If the rest is positive, the estimated coefficient will be greater than 1.

If the rest is negative but smaller in absolute value than cash flows, the coefficient will fall between 0 and 1.

If the rest is negative and larger in absolute value than cash flows, the coefficient will be negative.

In arithmetic terms, the value of the cash flow coefficient in the FHP model is not determined purely by an economic relationship but it is also a function of the omitted "rest." If the rest is systematically positive or negative across firms in a sample, it will predictably bias the coefficient away from its true economic value. If the ASI problem were non-existent, then segmenting the sample based on the sign of the identity's rest should have no significant effect on the results. If the researcher recalls the components of the rest, none of these omitted elements should influence, in theory, the estimated relationship between cash flows and investment.

With the aim of testing the measure of the ASI problem, a dichotomous variable representing a positive or negative rest (1/0) is constructed. Then I compute it as a multiplicative variable together with cash flows (ducf) and include it in equation [2].

$$Inv = a + b_1 * cf + b_2 ducf + \varepsilon \quad (3)$$

The logic of this test is as follows: The main hypothesis is that the databases in which I test the FHP model may be contaminated by the ASI problem, and therefore the coefficient of that multiplicative variable should be significant.

$H_1$: $b_2 \neq 0$

In case it is confirmed, this would indicate the presence of issues related to ASI. Recall that in the original Sanchez-Vidal (2023) article, the main model could still be correctly estimated in one of the estimations, but the significance of the multiplicative variable reveals the latent ASI problem. The mere presence of this bias is inherently problematic because it remains dormant until the dataset is stratified. For instance, subsampling based on variables correlated with the sign of the rest, such as the level of investment (Sánchez-Vidal, 2025) or the occurrence of an equity issuance, can activate the bias, rendering the resulting estimates unreliable.

An alternative to including a multiplicative variable would have been to partition the databases into subsamples based on the sign of the rest. However, the distinct advantage of incorporating a multiplicative variable is that it enables the analysis of the entire sample with a single equation estimation. An alternative specification could have included only the dummy variable for the sign of the rest. However, this approach would not isolate the specific distortion of the cash flow coefficient, which is the central parameter of interest in the FHP model. The multiplicative interaction term (dummy x cash flow) is essential for directly quantifying how the ASI problem biases this key estimate.

### 3. Model, variables and the set of previous databases

I will employ two models. The first is the *restricted model* (2), which is the standard formulation of the FHP (1988) model. The second is an *unrestricted model* (3) which augments the standard specification with a multiplicative variable (*ducf*), designed to capture the ASI effect.

Following the alternative specifications of the FHP (1988) model by Kaplan and Zingales (1997), the investment and cash flow variables are scaled by total assets. This approach was adopted for homogeneity, as the data required to calculate the replacement cost of the capital stock, as originally specified by FHP (1988), were unavailable for all databases. For the subsets where this calculation was feasible, I confirmed that the regression results were similar (not tabulated). Furthermore, to mitigate the influence of outliers, I trimmed the top and bottom 1% of both the investment and cash flow distributions

As previously noted, this analysis leverages a diverse collection of datasets to ensure robust and generalizable results. These datasets, associated with nine distinct published studies, vary across several dimensions: they encompass multiple countries and continents, cover different time periods spanning several decades, and comprise firms with heterogeneous characteristics. A detailed summary is provided in Appendix 1. All but one of the source articles are published in journals indexed in the ISI Web of Science. While several datasets originate from the same source database (SABI), the specific years, sampling frames, and firm subsets used are unique. The scale of the datasets also varies considerably, ranging from hundreds to hundreds of thousands of firm observations. To maximize the informational value, I utilized all available data points, even beyond what was used in the original publications. For instance, for the cross-sectional study labeled '2026-MD' in Table 1 of the following section, I incorporated multiple years of financial-accounting data that were available, rather than restricting the analysis to a single period.

## 4. Results and Discussion

Table 1 reports the estimation results for the restricted and unrestricted models across the nine databases. In addition to the estimated coefficients, their individual statistical significance, and the F-statistic for the overall model significance, I report the incremental fit provided by the unrestricted model. This improvement in explanatory power is computed in two ways: first, as the increase in the R² of the unrestricted model over the restricted model with $(R_U^2 - R_R^2)/R_U^2$, (Δ exp.power), which will show the percentage of total explained variance explained by recognizing the existence of the ASI problem. The $R^2$ employed in this formula corresponds to the unadjusted value, which is the one presented in the table. Secondly, I compute the **F$_{IF}$**, or the incremental fit, measured as: $F_{IF} = \frac{(RSS_R - RSS_U)/m}{RSS_U/(n-k)}$, where $RSS_R$ denotes the residual sum of squares from the restricted model, and $RSS_U$ represents the residual sum of squares from the unrestricted model. Here, m is the number of linear restrictions (or added parameters), n is the number of observations, and k is the number of variables in the unrestricted model (Gujarati, 1995). The computed F-statistic follows an F-distribution with degrees of freedom: $F \sim F_{m, n-k}$.

Given that the ASI reflects the growth of both sides of the balance sheet, I calculate the proportion of the increase in total assets attributable to investment (%inv/ΔTA) and the proportion of the increase in total funds attributable to cash flow (%cf/ΔTF). These metrics offer an informative glimpse into how closely the ASI approximates the full AI on each side of the balance sheet. The underlying full AI in the FHP model is based on the equality: ΔTotal Assets = ΔTotal Funds. It is important to recall that the short-term components of this identity, which are omitted from the standard FHP specification, are captured in Sánchez-Vidal (2023) within the rest variable, operationalized as the change in working capital, so that all components sum up the full AI.

The results in Table 1 demonstrate that the multiplicative variable (*dummy of rest x cf*) is statistically significant in all regressions and, notably, exhibits higher statistical significance than the standalone cash flow variable in every case. The improvement in model fit is substantial: the unrestricted model's explanatory power, as measured by R², is on average 8 to 9 times greater than that of the restricted model. On average, the unrestricted model accounts for 83.57% of the total explained variance. This dramatic increase in R² driven by the *ducf* variable is, however, artificial. It does not represent a genuine economic relationship but rather quantifies the statistical bias introduced by the ASI, as theorized by Sánchez-Vidal (2023). The magnitude of this R² inflation thus serves as a direct measure of the ASI problem's severity.

Table 1. Results of the FHP for the restricted and unrestricted models. The dependent variable is Investments, which is regressed on cash flows (cf). The unrestricted model includes a multiplicative variable of a dummy (1-positive rest of the ASI and 0-otherwise). The first F is the conventional overall F-test of the joint significance of the predictors of the model. I report a second measure of goodness of fit, such as the unadjusted $R^2$. RSS stands for the residual sum of squares. I report two indicators of the incremental fit of adding the multiplicative variable in the unrestricted model: the increase of explanatory power given by $(R_U^2 - R_R^2)/R_U^2$, and the $F_{IF}=\{(RSSR - RSSU)/m\}/\{(RSSU)/(n-k)\}$, and their respective significance, when compared to the critical values of a $F\sim F_{m,n-k}$. Finally, I report the percentage that the increase of investments and the cash flows represent of the increase in total assets. For the sake of brevity, I do not report the results for the intercept. [a] represents significance at the 1% level.

| | Coef | t | Coef | t | F | $R^2$ and Δ exp.power | RSS | $F_{IF}$ | %inv/Δ TA | %cf/ ΔTF |
|---|---|---|---|---|---|---|---|---|---|---|
| **Panel A: 2005-RQFA** | | | | | | | | | | |
| cf – (restricted) | 0.13 | 10.95 [a] | -0.05 | 4.90 [a] | 119.90 [a] | 1,60% | 61.500 | | | |
| ducf– (unrestricted) | | | 1.15 | 49.47 [a] | 1302.88 [a] | 25,50% | 46.519 | | | |
| % of full AI | | | | | | | | | 8.46% | 37.97% |
| Incremental fit | | | | | | 93,73% | | 2447,19 [a] | | |
| **Panel B: 2006-RFE** | | | | | | | | | | |
| cf – (restricted) | 0.12 | 28.29 [a] | 0.04 | 13.34 [a] | 800.38 [a] | 2.00% | 137.51 | | | |
| ducf– (unrestricted) | | | 1.34 | 154.6 [a] | 12606.62 [a] | 39.3% | 85.08 | | | |
| % of full AI | | | | | | | | | 3.86% | 41.30% |
| Incremental fit | | | | | | 94.91% | | 23982,3 [a] | | |
| **Panel C: 2008-** | | | | | | | | | | |
| cf – (restricted) | 0.04 | 7.73 [a] | -0.19 | -18.94 [a] | 59.73 [a] | 0.30% | 361.06 | | | |
| ducf– (unrestricted) | | [a] | 0.31 | 27.16 [a] | 399.64 [a] | 3.40% | 349,69 | | | |
| % of full AI | | | | | | | | | 7.37% | 39.12% |
| Incremental fit | | | | | | 91.18% | | 738,06 [a] | | |
| **Panel D: 2012-JBF** | | | | | | | | | | |
| cf – (restricted) | 0.06 | 28.73 | -0.08 | -34.79 [a] | 825.69 [a] | 0.50% | 1038.64 | | | |
| ducf– (unrestricted) | | | 0.64 | 137.5 [a] | 9926.28 [a] | 11.6% | 923.69 | | | |
| % of full AI | | | | | | | | | 3.77% | 21.24% |
| Incremental fit | | | | | | 95.69% | | 18926,7 [a] | | |
| **Panel E: 2014-EM** | | | | | | | | | | |
| cf – (restricted) | 0.10 | 23.77 [a] | -0.07 | -16.86 [a] | 565.18 [a] | 1.00% | 632.99 | | | |
| ducf– (unrestricted) | | | 76.27 | 95.07 [a] | 4842.43 [a] | 14.1% | 548.81 | | | |
| % of full AI | | | | | | | | | 4.32% | 15.91% |
| Incremental fit | | | | | | 92.91% | | 9061.61 [a] | | |
| **Panel F: 2022-FRL** | | | | | | | | | | |
| cf – (restricted) | 0.08 | 296.7 | -0.49 | -159.84 [a] | 88063.96 [a] | 1.32% | 59617.4 | | | |
| ducf– (unrestricted) | | | 0.69 | 883.4 [a] | 6326448 [a] | 12.00% | 52544.4 | | | |
| % of full AI | | | | | | | | | 0.00% | 11.43% |
| Incremental fit | | | | | | 89.00% | | 888941, [a] | | |
| **Panel G: 2021-IJFE** | | | | | | | | | | |
| cf – (restricted) | 0.01 | 0.08 | -0.22 | -3.21 [a] | 0.01 | 0.00% | 23.24 | | | |
| ducf– (unrestricted) | | | 1.44 | 10.43 [a] | 54.38% [a] | 13.5% | 20.11 | | | |
| % of full AI | | | | | | | | | 53.83% | 53.25% |
| Incremental fit | | | | | | 100.00% | | 108,48 [a] | | |
| **Panel H: 2022-JEB** | | | | | | | | | | |
| cf – (restricted) | 0.30 | 13.58 | 0.13 | 6.93 [a] | 184.35 [a] | 8.00% | 19.47 | | | |
| ducf– (unrestricted) | | | 1.30 | 31.28 [a] | 615.82 [a] | 37.80% | 13.04 | | | |
| % of full AI | | | | | | | | | 3,90% | 22.72% |
| Incremental fit | | | | | | 78.84% | | 20457,6 [a] | | |
| **Panel I: 2025-MD** | | | | | | | | | | |
| cf – (restricted) | 0.07 | 14.81 [a] | 0.00 | 0.07 | 219.27 [a] | 0.53% | 1085.76 | | | |
| ducf– (unrestricted) | | | 0.08 | 6.64 [a] | 131.78 [a] | 0.63% | 1084.60 | | | |
| % of full AI | | | | | | | | | 42.73% | 22.68% |
| Incremental fit | | | | | | 15.87% | | 44,37 [a] | | |
| **Total mean** | | | | | | 83.57% | | | 14.90% | 34.40% |

The results for *%inv/ΔTA* and *%cf/ΔTF* reveal that investments and cash flows often constitute a substantial portion of the full AI. In several cases, cash flows account for nearly the entire increase in total liabilities and equity, indicating that they are the primary driver of variation on the right-hand side of the balance sheet. While the proportion of total asset growth explained by investment appears smaller in some databases, this is attributable to the fact that the increase in total assets in those samples is primarily driven by changes in current assets (not tabulated), which is usually a more dynamic component.

The results indicate that the ASI problem is most severe in two scenarios: when one component of the identity constitutes a dominant proportion of the change on its side of the balance sheet (e.g., cash flows in Panel B), and second, when both sides contribute consistently to the growth in total assets (e.g., Panel G). However, significant bias can also occur with more modest percentages (e.g., Panel F). In all these cases, the arithmetic force of the identity overwhelmingly dominates, rendering the influence of true economic causality marginal.

The incremental fit provided by the duxcf variable is exceptionally large, so much so that it meets even the most stringent significance thresholds proposed in recent methodological research (e.g., Harvey, Liu, and Zhu, 2016). The F-statistic for the incremental fit not only surpasses the conventional 1% critical value in all databases but, in every case, exceeds the critical value for a probability of 0.0001% (not tabulated). This indicates a less than 1 in 1,000,000 probability that the observed improvement is due to random chance. The associated p-values are over 100,000 times more exigent than the stringent 1% confidence level. This result is consistent across all databases, including the 2026-MD sample (Panel I), which exhibited the lowest $R^2$ improvement. Consequently, the null hypothesis that the restricted model is valid can be decisively rejected.

These findings collectively demonstrate that the ASI problem is severe and systematically distorts the estimated coefficients in conventional FHP models. This is evident from the comparison of the cash flow coefficient between the restricted and unrestricted models, which exhibits erratic volatility, constantly changing in sign, magnitude, and statistical significance. Such instability is a known symptom of models based on near-complete AIs. As Felipe and McCombie (2009, p. 152) noted in the context of omitted variables or imperfect data aggregation, these conditions 'may, of course, lead to biases on the coefficients of the remaining variables, and sometimes in an unpredictable way.' The results provide robust empirical confirmation of this phenomenon, showing that the arithmetic force of the ASI fundamentally compromises the economic interpretation of the FHP model's parameters.

## 5. Open science and data reuse

In order to promote research transparency and verifiability the datasets supporting the findings have been deposited in a public repository. The databases are clearly described and available at:

https://figshare.com/s/bf3e56372333c789a524

In the link the researcher can find the 9 databases and the sintax for Stata to run the analyses. Given that restrictions apply to all databases as the data is property of the different providers, I provide only the transformed variables required for the analyses. These were computed from the original raw data, now removed, and all identifying information has been replaced with entirely new, unlinked identifiers. This is a way to ensure both reproducibility but also confidentiality, as the risk of reidentification is practically zero (Morehouse et al., 2024).

For researchers who wish to verify that the calculated variables are derived correctly from the original data, I offer access through a Secure Data Enclave. This is a controlled, remote-computing environment, analogous, for example, to the OECD Secure Data Enclave, where authorized researchers can analyze sensitive data without the ability to download, copy, or extract it (for access, please contact the corresponding author).

## 6. Conclusions

The model proposed by Fazzari, Hubbard, and Petersen (FHP, 1988), which aims to analyze the causal relationship between cash flows and investment, presents a fundamental methodological challenge. Contemporaneously, these variables are also linked by a linear accounting identity. Because the FHP model omits certain components of this full identity, it constitutes an Accounting Semi-Identity (ASI). This distinctive feature can introduce significant bias, as the estimated coefficients may be distorted by the "arithmetic force" of the incomplete identity.

To test whether results from the FHP model are biased by this ASI problem, I estimate an augmented specification that includes a multiplicative variable capturing the arithmetic influence of the omitted identity components. Using a diverse set of databases, varying in companies, countries, and time periods, I find the ASI issue to be both pervasive and severe. It is present in **100%** of the datasets analyzed. The explanatory power of the standard FHP model is minimal (approximately **16.5%** of the total explained variance), whereas the variable accounting for the ASI problem alone explains over **83%.** This indicates that the perceived "causal" relationship is largely spurious, dominated by what can be termed the "tyranny of the identity." Researchers might have previously assumed that, after acknowledging the FHP (1988) model as part of an accounting identity, any resulting bias in the estimated outcomes would be minimal, perhaps around 1%. However, this finding contradicts what was likely the prevailing assumption in the literature, an assumption that may help explain why the initial critique raised by Sánchez-Vidal (2023) remained largely unaddressed.

Given that this problem affects every database in the sample, the results suggest that a significant portion of prior empirical research published in top journals, with conclusions attributed to firm behavior, may instead reflect statistical artifacts arising from the necessity to satisfy the underlying accounting identity. This provides a compelling explanation for the numerous anomalous and counterintuitive results in the corporate finance literature concerning the FHP model, beginning with the influential critiques of Kaplan and Zingales (1997, 2000) till for example the more recent evidence by Lawrenz and Oberndorfer (2023) who even try to explain why the relation between cash flows and investments have declined and even turned negative in the recent years!. Future research must develop and adopt corrective strategies to mitigate ASI bias. Furthermore, to ensure the publication of sound, verifiable, and consolidated knowledge, journals should implement more pervasive and enforceable data-sharing policies, solidifying their role as a reliable reference within academia.

**Acknowledgements**
The authors would like to thank seminar participants at the University of Cartagena and the University of Calabria, for their helpful comments. In particular, the authors wish to specifically acknowledge Moncef Guizani, Bao Khac Quoc Nguyen, Bao Cong Nguyen To and Nham Thi Hong Nguyen for their exemplary commitment to data sharing. I extend my sincere thanks to Emanuele Bajo and Moritz Ritter, editors of the Journal of Economics and Business, for their genuine dedication to advancing Open Science, which sets a commendable standard in the field. This work was supported by 2023 Grant "Estancias de personal docente y/o investigador Senior en centros extranjeros" from the Spanish Ministerio de Universidades, with reference. PRX22/00694. All errors are my own.

| Table Appendix 1. Overview of the different databases used in this study. | |
|---|---|
| Panel A: 2005-RQFA | This database is provided by SABI from Informa, S.A. (associated with Bureau van Dijk), considering firms that have data for the entire sample period 1994–2000. The final sample is composed of 1,566 firms.<br>*Sample related to the published article:*<br>Sánchez-Vidal, J., & Martín-Ugedo, J. F. (2005). Financing preferences of Spanish firms: Evidence on the pecking order theory. Review of Quantitative Finance and Accounting, 25(4), 341-355. |
| Panel B: 2006-RFE | The final sample is taken from SABI from Informa, S.A. (associated with Bureau van Dijk), consisting of 4,865 companies for the 1993-2003 period.<br>*Sample related to the published article:*<br>Sánchez-Vidal, J., & Martín-Ugedo, J. F. (2006). Determinantes del conservadurismo financiero de las empresas españolas. Revista de economía financiera, 9, 47-66. |
| Panel C: 2008-SJFA | The final sample comes from the SABI database provided by Informa, S.A., covering the 1999-2002 period. It is a small sample because the companies need to have the auditing information for this period.<br>*Sample related to the published article:*<br>Duréndez Gómez-Guillamón, A. D., & Vidal, J. S. (2008). La influencia del informe de auditoría en la obtención de financiación bancaria. Spanish Journal of Finance and Accounting/Revista Española de Financiación y Contabilidad, 37(138), 255-278. |
| Panel D: 2012-JBF | The final sample is taken from AIDA database from Bureau van Dijk (BVD) and is composed of 17,165 firms having an average of 8.9 years per company, leading to an aggregate sample of 152,141 firm-year observations.<br>*Sample related to the published article:*<br>Bigelli, M., & Sánchez-Vidal, J. (2012). Cash holdings in private firms. Journal of banking & finance, 36(1), 26-35. |
| Panel E: 2014-EM | It is an incomplete data panel from the SABI database of Informa, S.A, consisting of 15,839 non-quoted companies for the 2001–2011 period. The final number of firm-year observations is 59.080.<br>*Sample related to the published article:*<br>Sánchez-Vidal, F. J. (2014). High debt companies' leverage determinants in Spain: A quantile regression approach. Economic Modelling, 36, 455-465. |
| Panel F: 2022-FRL | The study employs a large, unbalanced panel sample of 8.612.659 firm-year observations (collected from the AIDA database of Bureau van Dijk) of non-financial Italian companies over the 2012–2020 period.<br>*Sample related to the published article:*<br>Fasano, F., Sánchez-Vidal, F. J., & La Rocca, M. (2022). The role of government policies for Italian firms during the COVID-19 crisis. Finance Research Letters, 50, 103273. |
| Panel I: 2021-IJFE | The full sample consists of 1,110 firm-year observations from Saudi Arabia between 2004 and 2018.<br>*Sample related to the published article:*<br>Guizani, M. (2021). Macroeconomic conditions and i nvestment–cash flow sensitivity: Evidence from Saudi Arabia. International Journal of Finance & Economics, 26(3), 4277-4294. |
| Panel H: 2022-JEB | The sample consisted of Vietnamese listed firms with annual financial information collected from Thomson Reuters database and various sources, excluding financial firms as banks, insurance firms, funds, and firms with missing values, conforming an unbalanced panel data of 419 listed firms during 2010–2017, composed of 2115 firm-year observations.<br>*Sample related to the published article:* |
| | Nguyen, B. K. Q., To, B. C. N., & Nguyen, N. T. H. (2022). Unexpected money growth, nonfinancial firms as large shareholders and investment-cash flow relationship: Evidence from Vietnam. Journal of Economics and Business, 119, 106054. |
| Panel I: 2026-MD | The sample is a merger of an Orbis database by Bureau van Dijk (BVD), and the crossed information about managers by LexisNexis, and consists of 3,151 non-financial companies from 31 European countries for 2015.<br>*Sample related to the published article:*<br>La Rocca, M., Sanchez-Vidal, F.J. Fasano, F., & La Rocca, T. (2026). The Misconduct Ripple: How Corruption and Unethical Practices Affect Performance Across the Firm. Management Decision (forthcoming). |